September 23, 2012

# Unification of Gravity with Electromagnetism

Geoffrey F. Chew

*Theoretical Physics Group*
*Physics Division*
*Lawrence Berkeley National Laboratory*
*Berkeley, California 94720, U.S.A.*

Principles of Maxwell, Lorentz, Milne, Dirac and Feynman are combined to unify gravity with electromagnetism. Special-relativistic settled reality (*SR*) evolves, as universe age increases, via cosmological Feynman paths. Although *SR* is 'classical', its *evolution* is '*q*uantum mechanical'. A *unitary* Hilbert-space Lorentz-group representation--a *lightlike*-fiber-bundle—allows definition of *divergenceless* Lorentz-tensor self-adjoint *retarded-potential* operators. Feynman-path action (real but *not SR*) is invariant under a 7-parameter group that augments 6-parameter SL(2,c) by a 1-parameter compact (Kaluza-Klein) group generated by discrete electric charge. *Sublightlike* charged-particulate matter (a component of 'objective reality') reflects 'zitterbewegung'—*fluctuation* of a lightlike-velocity direction specified by 2 angles of a 3-sphere fiber. The third fiber angle is Dirac conjugate to a discrete-spectrum electric-charge operator, whose commutation with *all* potential operators renders *discrete* the spatially-localized temporally-stable *charged-particle* component of *SR*.

**Introduction**

Physics at molecular or smaller scales has relied on the special-relativistic 'particle' concept, but the *absence* of *finite*-dimensional *unitary* Lorentz-group representations precludes a Dirac-theoretic [1] Fock space that houses finite-spin 'Lorentz particles'. [The S matrix is *not* a 'Dirac theory'.] The present paper employs Gelfand-Naimark (*GN*) infinite-dimensional *Hilbert-space* unitary representation of the Lorentz-group [2] to define *classical retarded* gravitational-electromagnetic fields. A special-relativistic settled reality (*SR*) reconciles Dirac principles with 'detectable particles'. Particle properties such as electron mass are determined by an invariant Feynman-path action [3] that combines gravity with electromagnetism.

Deferring path action to another paper, we here *locally* represent (classical) gravity and electromagnetism by *expectations* of self-adjoint *retarded* Lorentz-*tensor* potential operators that 'Dirac-extend' the classical Lienard-Wiechert 4-*vector* retarded electromagnetic potential. Maxwell's equations are satisfied by *SR*.

Dalembertians of potential-operator expectations are current densities of conserved electric charge, energy and momentum. A 'detectable particle' is a discretely-charged *temporally-stable* 'concentration' in classical Lorentz-tensor densities. Our theory admits no *Dirac* meaning for 'particle wave function'. Although the Lemaître-Hubble redshift displayed by our proposal [4] might seem incompatible with special relativity, *all* our reality is Lorentz-group based.

The Gelfand-Naimark Hilbert space, that unitarily represents the 6-parameter semisimple 'Lorentz' Lie group SL(2,c), comprises normed twice-differentiable functions of location in a 6-dimensional super manifold—a fiber bundle of compact, SU(2), 3-sphere *fiber* over an invariantly metricized (noncompact) 3-hyperboloid *base space*. Astonishingly, this Hilbert space *also* represents a 12-parameter Lie group isomorphic to transformations of a unimodular 2×2 complex matrix through right *or* left multiplication by other such matrices.

The 6 generators of $SL(2,c)_R$ commute with the 6 generators of $SL(2,c)_L$. Referring to the product group as 12*L*, we here show how electric-charge *discreteness*, accompanying action symmetry under a 7-parameter 12*L subgroup*, underpins the experimental-physics (special-relativistic) meaning of 'individual particle'. Our ('cosmologically-relativistic') unified quantum theory of gravity and electromagnetism defines 'particle' to be a discretely-charged temporally-stable component of 'settled' (classical) reality.

Within each of 6 12*L* commuting-generator 'matched' pairs, one pair member generates displacements along a (noncompact) real line while the other generates displacements around a (compact) circle. Eigenvalues of the former are continuous while, of the latter, discrete. The action that determines universe Feynman-path (quantum) dynamics is invariant under both $SL(2,c)_R$ *and* a 1-parameter compact *left* subgroup of 12*L*. The *septet* of symmetry generators—*GN*-represented by 7 self-adjoint (Dirac) operators—are 'universe constants of motion'.

Central to this paper is a pair of *left* generators we call '*quc* energy' (continuous) and '*quc* electric charge' (discrete). The abbreviation '*quc*' stands not for reality but for a lightlike 'quantum-universe constituent' that, in collaboration with other *quc*s, Dirac-defines reality. An individual *quc*, despite lacking the temporal stability essential to objective reality, carries conserved momentum, angular momentum and electric charge in a Dirac wave-function sense. 'Dirac particle reality' is achieved through *expectations* of self-adjoint operators acting on a finite (although currently-huge) tensor product of *GN* single-*quc* Hilbert spaces.

Any *quc* momentum component—a right Lorentz-group generator in a *single-quc* Hilbert space-- commutes with the corresponding (same direction) component of the *quc*'s angular momentum. Two *different*-direction *quc* momentum components do *not* commute with each other. Although a corresponding property of *angular*-momentum operators is familiar in previous quantum theory, *quc*-momentum eigenvalues are continuous. *Quc*-angular-momentum eigenvalues (unsurprisingly) are discrete. Momentum operators generate displacements in the (non-compact) 3-dimensional metricized *hyperbolic quc*-fiber-bundle base space.

For pedagogical reasons we avoid the term 'Lorentz boost' that associates to a change of velocity. The velocity of a *quc* is unchangeably lightlike (always *c* in magnitude). On the other hand, *quc* momentum, despite being a 'boost generator', is a component of a 3-vector set of *quc spatial*-displacement generators within a generator 6-vector. Our *classical SR* meaning for *particle* momentum is more familiar; status as 3-vector component of a *positive* 4-vector there prevails.

A *fixed*-direction component of *quc* momentum generates an *infinitesimal* displacement in the *curved* base space. But, because of the curvature, a *non*-infinitesimal displacement in fixed direction does *not* follow a geodesic. In contrast, a lightlike displacement generated by the *quc*'s *energy* operator follows a geodesic that parallels *quc* velocity. *All* displacements of *GN* fiber-bundle *quc* location are at fixed 'age' within a 'big-bang' spacetime—the interior of a forward lightcone where redshift-specifying age is Minkowski distance from lightcone vertex. [4]

*Particle* momentum accompanies (settled) positive-energy objective-reality. A particle's temporal stability requires its energy to be positive. Probabilistic (Copenhagen) interpretation of the S matrix reflects 'hidden'--inaccessible to 'observation'--*negative*-energy reality. [4] *Another* paper, dealing with Feynman-path action, will define a classical *quc-path* momentum that, although *not* a *positive*-energy particle momentum, *is* a component of a right-Lorentz 4-vector.

*Quc-path* momentum *differs* from the 6-vector-component self-adjoint operator we call '*quc* momentum', while *both* these concepts differ from our reality's 4-vector '*particle* momentum'--conserved by vanishing divergence of gravitational potential rather than by action invariance. *Quc-path* 4-vector momentum (in contrast) is conserved by the *definition* of 'cosmological Feynman path' (*CFP*). Neither in ray nor in path is '*quc*' to be understood as 'particle'. ('Hidden reality', manifesting *negative*-energy *quc*s, is addressed in Reference (4).)

The universe spacetime representing the group 12*L* is a forward-lightcone interior. 'Age' enjoys a purely-classical 'arrowed' and scale-setting (redshift-specifying) status. Neither a group Casimir nor other nontrivial self-adjoint operator, age is invariant under the *full 12L* group. *Age,* 'arrowed' with unambiguous origin, plays a role similar in many ways to that of the *single* (classical-Newtonian, Euclidean-group-invariant) time of nonrelativistic (Heisenberg-Schrödinger-Dirac) quantum theory, despite the arbitrariness of that (non-arrowed) time's origin.

**Symmetry**

The group *12L* is isomorphic to transformations of a unimodular 2×2 complex matrix through left *or* right multiplication by some (other) such matrix. Any left *12L* element commutes with any right element (6-parameter left and right semisimple subgroups). *12L* invariance of a Haar measure accords this group a *unitary GN* Hilbert-space representation. [4]

Although all *12L generators* are represented by self-adjoint (Dirac) operators, not all *12L* elements correspond to symmetries. Path action is invariant under a 7-parameter symmetry subgroup, whose generator septet comprises the universe's 'constants of motion'. A left-Lorentz symmetry generator-- electric charge--joins the six right-Lorentz generators of momentum and angular momentum (a Lorentz 6-vector). Zero total-universe angular momentum corresponds to Mach's principle.

*GN* unitary 12*L* Hilbert-space representation is by functions of location in a 6-dimensional fiber-bundle at *fixed* age. Each bundle attaches a 3-sphere fiber to every location in a *12L*-invariantly metricized 3-dimensional hyperbolic base space that (independently of fiber) provides 4-vector representation of left *and* right Lorentz groups. Location in base space is '*quc* spatial location'. Two fiber angles specify '*quc*-velocity direction'; the third angle is Dirac (Kaluza-Klein) conjugate to *quc* electric charge. (No *quc* attributes, either in Hilbert space or in Feynman path, correspond to *SR*.)

**Definition of 'Reality' by *Expectation***

*Exceptional* ages, spaced by a Planck-scale age interval $\delta$, each carry a universe ray—a vector in a *quc* quasi-Fock space. Meaning for 'settled reality', within (*quc*-less although *quc-CFP-traversed)* slices of continuous 'classical spacetime' (where age is *not* exceptional), is through *classical* retarded right-Lorentz-tensor fields. These fields are prescribed by *expectations*, with respect to that *earlier* quasi-Fock-space ray whose age is closest to the field-point age, of self-adjoint multi-*quc* right-tensor retarded-field operators. The finite number of *qucs* doubles with each successive step of exceptional universe age. The minimum number-- that at universe beginning--was 1. Prominent among the classical fields prescribing settled local reality is a divergenceless second-rank symmetric (right) Lorentz-tensor-- *conserved energy-momentum current density*. This tensor field is the Dalembertian of our retarded gravitational potential, divided by Newton's gravitational constant.

We caution readers against attribution of 'ordinary-language' significance to the Hilbert-space term 'expectation'-- a real number prescribed by pairing some ray with a self-adjoint operator. *Absent* from our meaning for 'expectation' is any probabilistic or consciousness aspect. There is no 'Schrödinger cat'—only a 'cat'.

Electromagnetic 'expectation reality' has two distinct components, matching the unambiguous decomposition of the Lienard-Wiechert retarded vector potential into a zero-Dalembertian 'radiation' component proportional to electric-charge 'accelerations' and a 'Coulombic' component proportional to charge 'velocities'. (The Dalembertian of the latter is the conserved electric-charge current density.) 'Photon reality' (as recognized by experimental physicists) resides *partly* in the former vector-potential component *and* partly in the energy-momentum tensor.

**Unitary Hilbert-Space 12*L* Representation**

We now introduce the *GN* Hilbert-space *unitary* Lorentz-group representation. A later section will define the self-adjoint operators whose expectations prescribe reality. We invoke Pauli's 2×2 matrices, a tool familiar in particle theory although unexploited in Reference (2). A lightlike *single-quc* rigged Hilbert space provides 'regular' *12L* representation. Single-*quc* Hilbert-space vectors are twice-differentiable normed complex functions of the coordinates of a 6-dimensional ('super') manifold. Three (noncompact) dimensions spatially locate a *quc*, two (compact) specify its velocity direction and one (compact) underpins its charge. The latter dimension *also* distinguishes lightlike 'bosonic' *quc*s from 'fermionic'. (Analytic S-matrix CPT symmetry entangles 'internal' particle quantum numbers with the *complex* Lorentz group which preserves energy-momentum complex 4-vector inner products.)

Gelfand and Naimark defined a Hilbert space of functions of a unimodular 2×2 complex matrix $a$ through *three* complex variables $s$, $y$, $z$ (six real variables) according to the following product of three unimodular 2×2 matrices, each of which coordinates the manifold of a 2-parameter *abelian* 12 *L* subgroup:

$$a(s, y, z) = exp(-\sigma_3 s) \times exp(\sigma_+ y) \times exp(\sigma_- z). \qquad (1)$$

In Formula (1) the symbols $\sigma_1, \sigma_2, \sigma_3$ denote the standard (handed) set of Pauli hermitian traceless self-inverse 2×2 matrices (determinant −1), with $\sigma_\pm \equiv \frac{1}{2}(\sigma_1 \pm i\sigma_2)$. The matrix $\sigma_3$ is real diagonal while the real hermitian-conjugate matrix pair, $\sigma_+$ and $\sigma_-$, each has a unit off-diagonal element.

Three two-parameter subgroups are represented, respectively, by '*s*', '*y*' and '*z*' submanifolds of the 6-manifold $a$. The *z* manifold represents a 2-parameter 'velocity-direction' subgroup of *right* SL(2,c) while the *s* manifold represents the 2-parameter *left* subgroup of diagonal matrices that is central to this paper. The 2-dimensional *y* manifold, although associating neither to a right nor to a left subgroup, enjoys *geometrical* significance: A *directed* geodesic of Milne's hyperbolic 3-space [5] is specified by the pair *y*, *z* of complex variables.

The complex coordinate *z* specifies geodesic *direction* while *y* spatially locates the geodesic in a 2-dimensional surface *transverse* to this direction. Finally, a point *along* the *z*, *y* geodesic is *longitudinally* coordinated by *Re s*. Noteworthy is *absence* of any *geometrical* role for *Im s*; this absence is crucial to what follows.

Alternative to the unimodular-matrix factorization (1) is the factorization $a(s, y, z) = u(a) \times h(a_5)$, where $u(a)$ is *unitary* unimodular while $h(a_5)$ is *positive-hermitian* unimodular with $a_5 \equiv exp(-\sigma_3 Re\, s) \times exp(\sigma_+ y) \times exp(\sigma_- z)$. The matrix functions $u(a)$ and $h(a_5)$ we do not display here but are straightforwardly computable. The unitary $u(a)$ maps $a$ onto a compact 3-dimensional fiber space (a 3-sphere) that covers a noncompact *12L*-invariantly-metricized 3-dimensional base space onto which $a_5$ is mapped by $h(a_5)$. Base space will below alternatively be coordinated by a positive 4-vector of fixed invariant magnitude.

We employ Dirac's shorthand [1] of denoting, by a *single* symbol, *both* a (real classical) *quc* coordinate *and* a self adjoint *quc operator* whose spectrum comprises the possible values of this coordinate. An example is the symbol $Re\ s^\sigma$--linearly related to what might either be called 'the local time' of *Quc $\sigma$* or its 'longitudinal coordinate'. The symbol $E^\sigma$ will denote *Quc-$\sigma$* energy in (below-defined) '$\sigma$ local frame'. The 'canonically-conjugate' (when appropriately normalized) *Quc-$\sigma$* time and energy operators, $Re\ s^\sigma$ and $E^\sigma$, are below seen not to commute.

The 6-dimensional Haar measure,

$$d\boldsymbol{a}^\sigma = ds^\sigma\ dy^\sigma\ dz^\sigma, \tag{2}$$

is invariant under $\boldsymbol{a}^\sigma \to \boldsymbol{a}^{\sigma\Gamma} \equiv \boldsymbol{a}^\sigma \Gamma^{-1}$, with $\Gamma$ a 2×2 unimodular matrix representing a *right* SL(2,c) transformation of the coordinate $\boldsymbol{a}^\sigma$. The measure (2) is *also* invariant under analogous left transformation. The 'volume-element' symbol $d\xi$ in (2), with $\xi$ complex, means $d\ Re\ \xi \times d\ Im\ \xi$.

Any Hilbert-space *individual-quc* vector is a complex differentiable function $\psi(\boldsymbol{a}^\sigma)$ with invariant (finite) norm,

$$\int d\boldsymbol{a}^\sigma |\psi(\boldsymbol{a}^\sigma)|^2. \tag{3}$$

The integration in (3) spans the full $y^\sigma$ and $z^\sigma$ complex planes and the full $Re\ s^\sigma$ line but only a $2\pi$ interval of $Im\ s^\sigma$. Expectations of self-adjoint operators such as the *Quc-$\sigma$* discrete electric-charge,

$$Q^\sigma \equiv i\hbar^{1/2} g\ \partial/\partial Im\ s^\sigma, \tag{4}$$

with $g$ an elsewhere-addressed universal dimensionless constant, are specified by the norm (3).

The *Quc-$\sigma$* continuous-spectrum energy-operator companion to (4) is

$$E^\sigma_\tau \equiv i(\hbar/2\tau)\ \partial/\partial Re\ s^\sigma, \tag{5}$$

the symbol $\tau$ here standing for some *discrete* and *exceptiona*l value of universe age that labels a *quc* Hilbert space. (Our velocity unit is such that $c = 1$.) The six generators of (right) SL(2,c) are *also* self-adjoint linear homogeneous superpositions of partial first derivatives (in the $s,y,z$ basis)—derivative superpositions with coefficients dependent on $\boldsymbol{a}^\sigma$, but that *all* commute with (4) and (5) while *each* commutes with exactly *one* other member of the right-generator sextet.

Two positive-trace self-adjoint hermitian 2×2 unimodular-matrix functions, bilinear in $\boldsymbol{a}^\sigma$, $\boldsymbol{a}^{\sigma\dagger}$ and linear in $\tau$ that, together, are *equivalent* to the coordinate quintet $\boldsymbol{a}_5^\sigma$--$Re\ s^\sigma$, $y^\sigma$, $z^\sigma$--plus the age $\tau$ are

$$\boldsymbol{x}^\sigma(\boldsymbol{a}_5^\sigma, \tau) \equiv \tau\ \boldsymbol{a}^{\sigma\dagger}\boldsymbol{a}^\sigma, \tag{6}$$

$$\boldsymbol{v}^\sigma(\boldsymbol{a}_5^\sigma) \equiv \boldsymbol{a}^{\sigma\dagger}(\sigma_0 - \sigma_3)\ \boldsymbol{a}^\sigma. \tag{7}$$

The hermitian matrix $\boldsymbol{x}^\sigma(\boldsymbol{a}_5^\sigma, \tau)$ is a positive-timelike 4-vector with 'Lorentz magnitude' $\boldsymbol{x}^\sigma \bullet \boldsymbol{x}^\sigma = \tau^2$, while the hermitian matrix $\boldsymbol{v}^\sigma(\boldsymbol{a}_5^\sigma)$ is a positive-lightlike 4-vector ($\boldsymbol{v}^\sigma \bullet \boldsymbol{v}^\sigma = 0$) such that $x^\sigma \cdot v^\sigma = \tau$. The dimensionful 4-vector $x^\sigma$ locates *Quc* $\sigma$ in spacetime (with respect to lightcone vertex) while dimensionless $v^\sigma$ specifies this *quc*'s lightlike velocity 4-vector--whose timelike component equals 1 in the *Quc-σ local frame* where $x^\sigma = (\tau,0,0,0)$.

Our definition of 'reality' supposes present-universe age $\tau$ to be huge compared to an ('inflation'-interpretable) 'starting-age' $\tau_0$ that in turn was huge compared to the Planck-scale exceptional-age spacing $\delta$. The universe starting age $\tau_0$ establishes the 'macro' scale of laboratory physics (~ 1 km)--in effect determining Avogadro's number. Physics (as pursued by humanity) recognizes a unique right-Lorentz local frame--the frame in which cosmic background radiation is isotropic, reflecting initial-universe spherical symmetry.

Failure of the commutator,

$$[E^\sigma(\tau), v^\sigma(a_5^\sigma)] = i\hbar/\tau \, v^\sigma(a_5^\sigma), \tag{8}$$

to vanish at finite $\tau$ reflects redshift.[4]

One might suppose '*quc* lightlikeness' to preclude massive particles, but we shall define reality via *expectations* of self-adjoint lightlike gravitational and electromagnetic potential *operators* that are functions of the (right) 4-vector *Quc-σ* operators $x^\sigma(a_5^\sigma, \tau), v^\sigma(a_5^\sigma)$ and the invariant operators $E^\sigma(\tau)$ and $Q^\sigma$. *Fluctuation* of *quc* velocity-direction, at *fixed* momentum, helicity, energy and electric charge, is expected in the universe wave function. Conserved ('classical') current densities of energy-momentum and electric-charge are then generally *not* 'lightlike'. Although Dirac's lightlike electron-velocity operator, which led Schrödinger to coin the term, 'zitterbewegung', differs importantly from the *quc*-velocity operator (7), Schrödinger's language is appropriate for describing 'cosmological origin of rest mass'.

**Classical Retarded Settled Reality**

We now blend the *classical* Lienard-Wiechert (*LW*) *retarded-field* notion with *fixed-age* $\tau = N\delta$ Hilbert-space self-adjoint *quc* operators. The reader is warned against confusing the self-adjoint field operators defined here with the *quantum radiation fields* employed by the Standard Model-- the latter operators *not* being Dirac-associable to *SR*.

Let the positive-timelike 4-vector symbol $x$, with $(N\delta)^2 < x \cdot x < [(N+1)\delta]^2$, denote a spacetime location *between* the ages $N\delta$ and $(N+1)\delta$. We shall define a continuous set of $x$-labeled single-*quc* self-adjoint operators on the $N$ (exceptional-age) Hilbert space housed at Age $N\delta$. Summing over *all qucs*, settled reality is then prescribed by interpreting a corresponding continuum of Ray $N$ expectations as the divergenceless (retarded, classical) *LW* electromagnetic vector potential. First derivatives of this vector potential yield Maxwell's electric and magnetic fields; the potential's Dalembertian is the conserved electric-charge current density. Maxwell's (classical) equations, for electric and magnetic fields in terms of current density, apply throughout the interior of the spacetime slice.

In what follows the single superscript $\sigma$ is to be understood as identifying *Quc* $\sigma$ at age $N\delta$. For the retarded electromagnetic vector-potential operator $A^\sigma_\mu(x)$, associated to *Quc* $\sigma$ and the spacetime-location $x$ (the 'field-point'), we postulate

$$A^\sigma_\mu(x) \equiv \theta_{\text{ret}}(x, a_5^\sigma) \, Q^\sigma \, v^\sigma_\mu / v^\sigma \cdot (x - x^\sigma), \tag{9}$$

the retardation step function $\theta_{ret}(\boldsymbol{x}, \boldsymbol{a}_5{}^\sigma)$--defined two paragraphs below--*not* depending on *Im s$^\sigma$*. All operators in (9) commute. The (right) 4-vector operators $x^\sigma$ and $v^\sigma$ have been defined, respectively, by Formulas (6) and (7). Because the *quc*-velocity 4-vector $v^\sigma$ is lightlike, the Lorentz-divergence of $A^\sigma{}_\mu(\boldsymbol{x})$ vanishes—a consideration that will render electric-charge conservation an aspect of 'classical reality'.

The $\boldsymbol{x}$ dependence of the potential is seen to reside in the step function $\theta_{ret}(\boldsymbol{x}, \boldsymbol{a}_5{}^\sigma)$ and in the invariant *LW*-denominator operator, $v^\sigma \cdot (\boldsymbol{x} - \boldsymbol{x}^\sigma)$. Because $v^\sigma \cdot v^\sigma = 0$, this denominator has the same value at *all quc* spacetime locations (not only that of age $\mathcal{N}\delta$) along the lightlike trajectory with *Quc-$\sigma$* velocity that passes through $\boldsymbol{x}^\sigma$. If the $\boldsymbol{a}_5{}^\sigma$ trajectory *intersects* the $\boldsymbol{x}$ backward lightcone, *classical LW* language refers to that intersection's location as *the* spacetime location of the 'retarded source' for the electromagnetic potential $A^\sigma{}_\mu(\boldsymbol{x})$. Age $\mathcal{N}\delta$ *quc*s whose *spatial* locations are *far* from that of the slice-interior point $\boldsymbol{x}$, thereby admit *LW* description as being located in the 'distant past' of $\boldsymbol{x}$.

The symbol $\theta_{ret}(\boldsymbol{x}, \boldsymbol{a}_5{}^\sigma)$ in (9) denotes a function equal to 1 *iff* the $\boldsymbol{a}_5{}^\sigma$ trajectory (passing with velocity $v^\sigma$ through the *Quc-$\sigma$* spacetime location $\boldsymbol{x}^\sigma$) intersects the $\boldsymbol{x}$ *backward* lightcone. Otherwise $\theta_{ret}(\boldsymbol{x}, \boldsymbol{a}_5{}^\sigma)$ vanishes. (*Any* lightlike trajectory not located *on* the $\boldsymbol{x}$ lightcone intersects the $\boldsymbol{x}$ forward-backward lightcone exactly *once*.) Summed over all sources, the Ray-$\mathcal{N}$ expectation of (9) prescribes the classical electromagnetic vector potential $A^{\mathcal{N}}{}_\mu(\boldsymbol{x})$ within the Ray-$\mathcal{N}$ immediate future.

Although the symbol $A_\mu(\boldsymbol{x})$, *without* superscript $\mathcal{N}$, may be employed to designate the retarded vector potential *almost everywhere* in spacetime, exclusion must be remembered of the exceptional ray ages where $\tau = \mathcal{N}\delta$. Classical reality is not defined *on* the hyperboloids that house rays. Second-order differential equations connecting (classical) potentials to current densities, that are meaningful *inside* any 'spacetime slice', only *approximately* extrapolate these fields from one slice to the next. Universe reality evolution is *quantum-mechanically* determined by cosmological Feynman paths (*CFP*s) --via action-specified *phases* of *complex* unimodular numbers. [*CFP* prescription is *not* addressed here.]

We now attend to a settled *gravitational* reality founded on the self-adjoint *quc* energy operator (5) rather than the *quc* electric-charge operator (4). The corresponding pair of commuting left-Lorentz-group generators, invariant under the right Lorentz group and representable by individual *qucs*, are Dirac conjugate to real and imaginary parts of the complex *quc* coordinate $s^\sigma$.

Paralleling the electromagnetic divergenceless vector potential $A_\mu(\boldsymbol{x})$ is a gravitational divergenceless symmetric-tensor potential $\Phi_{\mu\nu}(\boldsymbol{x})$. When divided by $G$, the Dalembertian of $\Phi_{\mu\nu}(\boldsymbol{x})$ prescribes (without Heisenberg uncertainty) the 4-vector current density of conserved energy-momentum. We anticipate qualitative difference between electromagnetic and gravitational objective reality, symptomized by positivity of *Maxwell-field* energy density. We expect to categorize *all* electromagnetic reality as 'objective' whereas gravitational reality comprises not only objective but also 'hidden' negative-energy components that have necessitated a probabilistic interpretation for Copenhagen quantum theory.

The (Newton-*LW*-Dirac) gravitational-potential operator

$$\Phi^\sigma{}_{\mu\nu}(\boldsymbol{x}) = G\,[E^\sigma V^\sigma{}_{\mu\nu}(\boldsymbol{x}) + V^\sigma{}_{\mu\nu}(\boldsymbol{x}) E^\sigma], \tag{10}$$

where the divergence-less symmetric-tensor retarded-field self-adjoint operator $V^\sigma{}_{\mu\nu}(\boldsymbol{x})$, defined by

$$V^\sigma{}_{\mu\nu}(\boldsymbol{x}) \equiv \theta_{ret}(\boldsymbol{x}, \boldsymbol{a}_5{}^\sigma)\, v^\sigma{}_\mu v^\sigma{}_\nu / v^\sigma \cdot (\boldsymbol{x} - \boldsymbol{x}^\sigma), \tag{11}$$

*right* transforms as a symmetric second-rank Lorentz tensor of zero invariant trace. Paralleling the electromagnetic vector potential, the tensor classical gravitational potential $\Phi^{\mathcal{N}}{}_{\mu\nu}(\boldsymbol{x})$ is the Ray $\mathcal{N}$ expectation of (10), summed over all *quc*s (whose *total* number is finite). The Dalembertian of $\Phi^{\mathcal{N}}{}_{\mu\nu}(\boldsymbol{x})$, when divided by $G$, is the energy-momentum tensor --presumed to manifest both massless and massive particles (e.g., photons, electrons, atoms) *and* electromagnetic-field energy-momentum, all positive-energy components of objective reality. 'Hidden reality', associated to negative energy but also manifested by $\Phi_{\mu\nu}(\mathbf{x})$, is elsewhere addressed. [4]

**Conclusion**

We have unified gravity and electromagnetism through a succession of *quc* quasi-Fock-space rays that unitarily represent a 12-parameter 'right-left doubling' of SL(2,c). Self-adjoint *quc* energy (source of gravity) and *quc* electric charge (source of electromagnetism) are, respectively, Dirac conjugate to real and imaginary parts of a complex Gelfand-Naimark coordinate for a 6-dimensional (super) fiber-bundle manifold. Sustained has been 'Dirac reality'-- via self-adjoint Hilbert-space operators--as well as principles of Maxwell and Feynman.

A 7-parameter symmetry group (with a 6-parameter Lorentz subgroup) is generated by 7 conserved self-adjoint operators— momentum, angular momentum and electric charge —whose expectations separately aggregate to zero for the universe as a whole. Elsewhere described is a *single-quc* spherically-symmetric *initial* condition that specifies 'zero total-universe energy'. Classical 'settled' reality resides in electromagnetic and gravitational fields within invariant spacetime slices of Planck-scale age width. Each universe ray is separated from its successor by such a slice.

Any ray (after the first at Age $\tau_0$) is determined from its predecessor by the actions of cosmological Feynman paths that traverse the intervening slice. Gravitational and electromagnetic line-integral path action, proportional to the potentials here defined, is specified in a separate paper. Although no spacetime-slice-inhabiting particle (an *SR* component for some age interval where $\tau \gg \tau_0$) lives forever, a *quc* never dies. The number of *quc*s at age $\mathcal{N}\delta$ -- equalling 2 raised to the power, $\mathcal{N} - \mathcal{N}_0$, where $\mathcal{N}_0 = \tau_0/\delta$, increases monotonically with universe age.

**Acknowledgements**

Intense discussions over two decades with Henry Stapp are reflected in this paper. Help has also been received from Jerry Finkelstein, David Finkelstein, Dave Jackson, Don Lichtenberg, Stanley Mandelstam, Ivan Muzinich, Ralph Pred, Ramamurti Shankar, Eyvind Wichmann and Bruno Zumino.